\documentclass[sn-mathphys]{sn-jnl}

\usepackage{amsthm,amsmath}

\usepackage{amsfonts}
\usepackage{amsmath,verbatim,color,amssymb,epsfig,bm,
mathrsfs,amsthm,latexsym,subfigure,graphicx}
\usepackage[numbers]{natbib}
\usepackage{hyperref}
\hypersetup{colorlinks=true,linkcolor=blue,citecolor=blue,urlcolor=blue}
\usepackage{array}
\usepackage{graphics}
\usepackage{subfigure}
\usepackage{multirow}

\usepackage{appendix}

\usepackage[figuresright]{rotating}

\usepackage{url}

\usepackage{nameref}

\usepackage{amsmath,verbatim,color,amssymb,epsfig}
\usepackage{graphicx}
\usepackage{threeparttable}

\jyear{2021}%

\theoremstyle{thmstyleone}%
\newtheorem{theorem}{Theorem}
\newtheorem{proposition}[theorem]{Proposition}%

\theoremstyle{thmstyletwo}%
\newtheorem{example}{Example}%
\newtheorem{remark}{Remark}%

\theoremstyle{thmstylethree}%
\newtheorem{definition}{Definition}%

\begin{document}

\makeatletter 
\@addtoreset{equation}{section}
\makeatother  

\def\yincomment#1{\vskip 2mm\boxit{\vskip 2mm{\color{red}\bf#1} {\color{blue}\bf --Yin\vskip 2mm}}\vskip 2mm}
\def\squarebox#1{\hbox to #1{\hfill\vbox to #1{\vfill}}}
\def\boxit#1{\vbox{\hrule\hbox{\vrule\kern6pt
          \vbox{\kern6pt#1\kern6pt}\kern6pt\vrule}\hrule}}

\def\theequation{\thesection.\arabic{equation}}
\newcommand{\ds}{\displaystyle}

\newcommand{\bcU}{\boldsymbol{\cal U}}
\newcommand{\bbeta}{\boldsymbol{\beta}}
\newcommand{\bdelta}{\boldsymbol{\delta}}
\newcommand{\bDelta}{\boldsymbol{\Delta}}
\newcommand{\boldeta}{\boldsymbol{\eta}}
\newcommand{\bxi}{\boldsymbol{\xi}}
\newcommand{\bfeta}{\boldsymbol{\eta}}
\newcommand{\bGamma}{\boldsymbol{\Gamma}}
\newcommand{\bSigma}{\boldsymbol{\Sigma}}
\newcommand{\balpha}{\boldsymbol{\alpha}}
\newcommand{\bOmega}{\boldsymbol{\Omega}}
\newcommand{\btheta}{\boldsymbol{\theta}}
\newcommand{\bmu}{\boldsymbol{\mu}}
\newcommand{\bnu}{\boldsymbol{\nu}}
\newcommand{\bgamma}{\boldsymbol{\gamma}}
\newcommand{\bpsi}{\boldsymbol{\psi}}
\newcommand{\bphi}{\boldsymbol{\phi}}
\newcommand{\bomega}{\boldsymbol{\omega}}
\newcommand{\newpar}{{\vspace{0.15cm} \noindent}}
\newcommand{\indep}{\mbox{$\,\perp\!\!\!\perp\,$}}
\newcommand{\nindep}{\mbox{$\,\not\!\perp\!\!\!\perp\,$}}
\newcommand{\toadd}{\vspace{1.5cm} \begin{center}
{\LARGE \color{red} to add} \end{center} \vspace{1.5cm}}

\newcommand{\comm}[1]{}


\setcounter{secnumdepth}{3}


\title[Bayesian Mendelian randomization with interval null hypotheses]{Bayesian Mendelian randomization testing of interval causal null hypotheses: ternary decision rules and loss function calibration.}


\author[1]{\fnm{Linyi} \sur{Zou}}

\author[2]{\fnm{Teresa} \sur{Fazia}}

\author[1]{\fnm{Hui} \sur{Guo}}
\equalcont{Joint senior authors}

\author*[1]{\fnm{Carlo} \sur{Berzuini}}\email{carlo.berzuini@manchester.ac.uk}
\equalcont{Joint senior authors}

\affil*[1]{\orgdiv{Centre for Biostatistics, School of Health Sciences}, \orgname{The University of Manchester}, \orgaddress{\street{Jean McFarlane Building, Oxford Road}, \city{Manchester}, \postcode{M13 9PL}, \country{UK}}}

\affil[2]{\orgdiv{Department of Brain and Behavioural Sciences}, \orgname{University of Pavia}, \orgaddress{\city{Pavia}, \postcode{27100}, \country{Italy}}}





\abstract{We enhance the Bayesian Mendelian
Randomization (MR) analysis framework
of \citet{Carlo2018} by incorporating a
novel Bayesian method for testing
interval null causal hypotheses, in the
interest of a healthier approach
to causal discovery. A number of ideas are combined in our
proposal. First, we
replace the usual point null hypothesis for the causal effect
with a region of practical equivalence (ROPE), in the spirit of
\citet{Kruschke2018}. Second, we allow the hypothesis
test decision to be taken on the basis of the Bayesian posterior odds
of the causal effect falling within the ROPE, calculated via
a new "Markov chain Monte Carlo cum Importance sampling" procedure.
Third, we allow the test decision rule to accommodate
an {\em uncertain} outcome, for those situations where
the posterior odds is neither large nor small enough. Finally,
we present an approach to calibration of the proposed
method via loss function. We illustrate the method with
the aid of a study of the causal effect of obesity on
risk of juvenile myocardial infarction based on a unique
prospective dataset.}




\maketitle

\section{Introduction}\label{sec1}
\indent

The causal effect of a modifiable exposure on an outcome can, under
certain assumptions, be assessed from observational data by
using measured variation in genes as an instrumental variable.
This is called a Mendelian Randomization (MR) analysis
(\citet{Martjin1986}; \citet{George2003}; \citet{Debbie2008};
\citet{wooldridge2009}). Standard approaches to MR
assume a parametric data generating model where the
unknown magnitude of the causal effect of interest
is represented by a parameter $\beta$ say. At least initially, we
assume $\beta$ to be a scalar.
In these approaches the hypothesis of a
null causal effect, $H_0$, is commonly defined as $\beta$
taking value 0. This is referred to as a {\em point null}
hypothesis: $H_0 : \beta=0$. In which case the alternative hypothesis
is $H_1: \beta \ne 0$. Whenever $H_0$ is rejected,
a "discovery" is claimed.

\vspace{0.1cm}

An alternative approach called {\em interval null}
hypothesis testing defines the causal null hypothesis
as $H_0: \mid\mid \beta \mid\mid \; \le T $,
which implies $H_1: \mid\mid \beta \mid\mid \; \gt T $, where a positive $T$
is specified by the user in such a way that $[-T, T ]$ represents an
interval of causal effect values that are practically equivalent
to zero. The $[-T, T]$ interval may in fact be regarded as an
example of {\em Region of Practical Equivalence} (ROPE),
in the sense of \citet{Kruschke2018}
In the present context, one advantage of the ROPE formulation
is that it tends to prevent the null from being
rejected in favour of a minuscule causal effect
that is a pure consequence
of an inevitably "imperfect" model being applied to a large amount of data.
The ROPE approach is discussed by a number of authors, including \citet{Jeffrey2020};
\citet{RikoK2021}; \citet{Riko2021}; \citet{Liao2021};
\citet{Linde2021} and \citet{Nathaniel2022}.

\vspace{0.1cm}

In both the mentioned situations, the causal null test
can be performed in a frequentist or in a Bayesian way.
One drawback of the frequentist approach is that the
test decision has to be made without
quantifying the relative amounts of evidence
in favour of $H_0$ and $H_1$ (\citet{JamesO1987}). This
is a major motivation for adopting a Bayesian approach to
the testing.

\vspace{0.1cm}

For example in the ROPE approach of \citet{Kruschke2018}, the user
specifies a Bayesian prior on $\beta$ over the entire admissible
space for this parameter, so that a Bayesian
posterior distribution for $\beta$ can be calculated.
Then the test decision is made by looking at the position
of the credible
interval for $\beta$ relative to the ROPE. In particular,
whenever the ROPE and the credible interval overlap,
the user may be willing to declare an "uncertain" outcome. One drawback
of this approach is the high
sensitivity of the conclusion to the choice of $T$.

\vspace{0.1cm}

One of our present aims is to introduce the interval null
approach to causal hypothesis
testing within the
Bayesian MR framework of \citet{Carlo2018},
further refined by  \citet{Zou2020}, so as to avoid the
limitations of frequentist testing, as well as excessive sensitivity to
ROPE specification.

\vspace{0.1cm}

Our proposed method involves the specification of a flat prior for $\beta$
over the ROPE under $H_0$, and a vague prior for $\beta$ outside the
ROPE under $H_1$, after which we use a novel method that
combines Markov chain Monte Carlo (MCMC) and
Importance sampling technology to calculate the posterior odds
for $\beta$ falling within the ROPE, which will then determine
the outcome of the test. In particular, in those situations where
the posterior odds is neither large nor small enough, indicating
insufficient evidence in favour of either hypothesis,
the test decision may be declared "uncertain", which represents a move
from a binary to a ternary decision logic.

\vspace{0.1cm}

Our proposed approach reduces the risk
of an MR analysis leading to
a causal discovery claim in the presence of only scant evidence in favour of the
alternative, as well as the risk of the null hypothesis being
accepted in spite of the data being compatible with
a (possibly important) causal effect. A further advantage is the reduction of
the risk of artefactual discoveries where an estimated causal effect of negligible
magnitude appears to be significant only as a consequence
of discrepancies between data and model -- a highly likely phenomenon in MR analysis.

\vspace{0.1cm}

Finally, for purposes of
decision rule calibration,
we introduce a loss function
that measures the cost incurred by each possible combination of
a decision outcome and a value for the true causal effect. We
illustrate this method for model parameter tuning and comparison
of a classical and a Bayesian approach to MR.

\vspace{0.1cm}

We illustrate our proposed method with the aid of a MR study
of the effect of obesity on risk of juvenile myocardial infarction,
on the basis of a unique set of data from patients hospitalized for myocardial
infarction and healthy individuals aged between 40 an 45 years in Italy.

\section{Methods}
\indent


\subsection{Bayesian Mendelian Randomization Model}
\indent

Suppose we wish to assess the putative causal effect of a scalar exposure $X$ on a scalar outcome $Y$, by using information provided by a set $\mathbf{Z} \equiv (Z_1, \ldots , Z_J)$ of instrumental variables (IVs, or instruments), typically genetic variants. A directed acyclic graph (DAG) representation of the proposed model for this task is shown in Figure \ref{figure1}.

Suppose, for the time being, that each individual in the sample comes with a completely observed set of variables $(X, Y, \mathbf{Z})$. Without infringing the argument's general validity, let $Y$ be a binary variable. Let $U$ denote a scalar summary of the unobserved confounders of the relationship between $X$ and $Y$. Within a Bayesian framework, if we assume standardised $(\mathbf{Z}, X)$ variables and linear additive dependencies, then a possible -- fully identifiable -- parametrization of the model is:

\vspace{-0.3cm}

\begin{figure}[h!]

\centering

\includegraphics[scale=0.3]{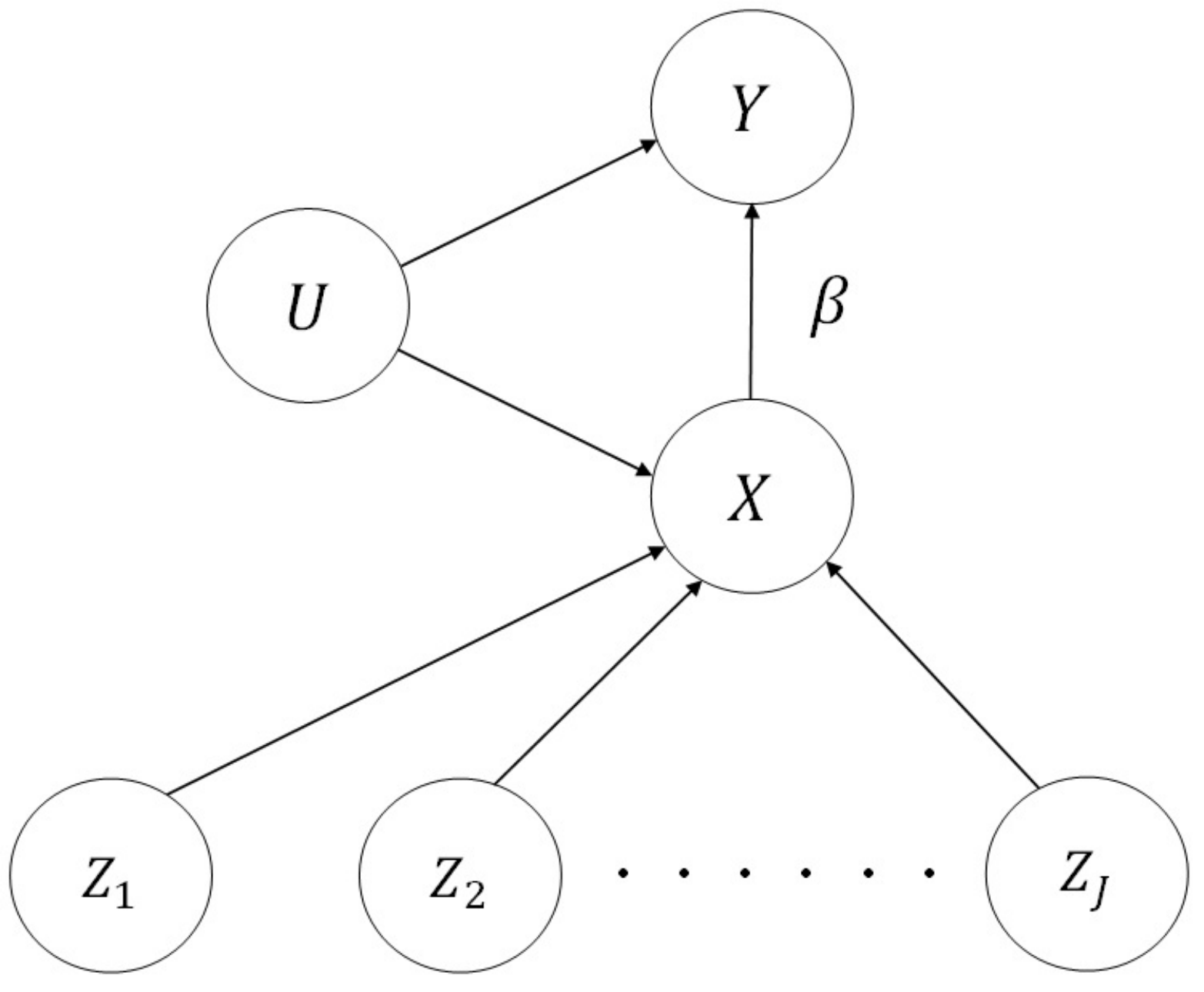}\\

\caption{Directed acyclic graph (DAG) representation of the Mendelian randomisation model we consider throughout the paper. The $X \rightarrow Y$ arrow is represented in the model equations by unknown parameter $\beta$.}

\label{figure1}

\end{figure}

\vspace{-1cm}

\begin{align}
  U &\sim N(0, 0.1), \label{2.1}\\
  X \mid \mathbf{Z},U &\sim N\left(\sum\limits_{k=1}^{J}\alpha_{k}Z_{k}+\delta_{X}U, \sigma_{X}^{2}\right), \label{2.2}\\
  \mu &= \text{expit}(\omega+\beta X + \delta_{Y}U), \label{2.3}\\
  \rule{0cm}{0.7cm}Y &\sim \text{Bernoulli}(\mu), \label{2.4}
\end{align}
where the symbol $\sim$ stands for ``is distributed as'' and $\text{expit}(a) \equiv \frac{e^a}{1+e^a}$. The symbol $N(a,b)$ denotes normal distribution with mean $a$ and variance $b$, and $\sigma_X$ is the standard deviation of an independent random perturbation of $X$. Of inferential interest is the causal effect of the exposure $X$ on the outcome $Y$, as quantified by parameter $\beta$, with $\beta=0$ corresponding to {\em absence} of the causal effect.
The vector parameter $\balpha = (\alpha_{1}, \alpha_{2}, \ldots, \alpha_{J})$ represents the strengths of the pairwise associations between instruments and exposure.
This model is adapted from  \citet{Carlo2018} and \citet{Zou2020}.

As is common in statistics, the DAG can be interpreted as a way of coding conditional
independence relationships that are implicit in the model equations.
These relationships are conveniently expressed by using the conditional independence notation of \citet{Dawid1979}, where
$A \indep B \mid C$ reads: ``$A$ is independent of $B$, given $C$'', asserting that the conditional distribution of the random variable $A$, given the value of the random variable $C$, does not further depend on $B$. Note that $A$ is marginal independent of $B$ when $C$ is empty, denoted as $A \indep B$.
Our model equations
and Figure \ref{figure1} are consistent with the following two conditions:

\begin{enumerate}

\item $\mathbf{Z} \indep U$: confounder independence

\item $Y \indep \mathbf{Z} \mid (X,U)$: exclusion-restriction.

\end{enumerate}

Condition 1 states that
each of the instrumental
variables included
in $\mathbf{Z}$ is independent of $U$. This
condition is untestable, due
to the fact that $U$
is unobserved.

Condition 2 states that there is no association between $\mathbf{Z}$ and $Y$ other than that mediated by $X$, and can be at best only partially tested. An additional condition is that association between each IV in $\mathbf{Z}$ and $X$ is not null.

Prior specifications required to complete the Bayesian formulation of the model were discussed at length in \citet{Carlo2018}. In our simulations, we have taken $\sigma_{X}$ to follow a priori an inverse-gamma distribution, $\sigma_{X} \sim$ \emph{Inv-Gamma}(3, 2), and each component of $\balpha$ to be independently normally distributed with mean 0.5 and standard deviation 0.2:
\begin{equation*}
  \balpha =  \left(
      \begin{array}{c}
        \alpha_{1} \\
        \alpha_{2} \\
        \vdots \\
        \alpha_{J} \\
      \end{array}
    \right) \sim N\left[\left(
                     \begin{array}{c}
                        0.5 \\
                        0.5 \\
                        \vdots \\
                        0.5 \\
                     \end{array}
                 \right),\left(
                            \begin{array}{cccc}
                                 {0.2}^{2} & 0 & \cdots & 0 \\
                                 0 & {0.2}^{2} & \cdots & 0 \\
                                 \vdots & \vdots & \ddots & \vdots \\
                                 0 & 0 & \cdots & {0.2}^{2} \\
                            \end{array}
                         \right)\right].
\end{equation*}
The prior for $\beta$ will be discussed later in this paper.

\vspace{0.3cm}

\subsection{Region of Practical Equivalence (ROPE)}

\indent

Having specified the model equations (\ref{2.1}-\ref{2.4}), one would often let the null causal hypothesis
 (asserting that the causal effect is non-existent)
 be defined by $\beta = 0$.
This choice raises a problem: the subspace of data generating processes corresponding to a ``non-existent'' causal effect is singular with respect to the full space of data generating processes
defined by the model,
so that, for a continuous
prior for $\beta$, the posterior probability of
a nonexistent causal effect is zero.
\color{black} From a practical point of view, when inference computations are performed via Markov chain Monte Carlo (MCMC) simulation (\citet{Nicholas1953}) with a continuous prior on $\beta$, the probability of the chain visiting the point-space
 $\beta=0$ is zero, which prevents us from calculating
and comparing posterior probabilities for the ``non-existence'' and for ``existence'' hypotheses.
This explains why Bayesians resort to Bayes factors. The reason is that Bayes factors avoid the problem by
comparing marginal likelihoods, albeit incurring other practical and conceptual difficulties.
A popular (albeit risky) option is to treat Bayesian posterior credible intervals as if they were classical confidence intervals, that is, by deciding in favour of the null (non existence) if and only if the credible interval covers the value 0. This option leads to generally incorrect inferences, thereby acting against scientific reproducibility (and against a fair comparison of Bayesian vs classical methods).

\vspace{0.2cm}

We avoid the singularity problem by defining ``non existence'' of the causal effect as corresponding to the value of $\beta$ falling within a {\em neighborhood} around $\beta=0$, called the {\em region of practical equivalence} (\citet{Kruschke2018}), or the {\em region of practical importance}, hereafter referred to as ROPE. This enables us to calculate a posterior probability over the ``non-existence'' subspace, as a basis for the hypothesis test decision. We shall claim a ``discovery'' when the posterior probability of $\beta$ falling in the ROPE does not exceed a specified threshold. For a user-specified real positive $T$, the non-existence (of the causal effect) and existence hypotheses, $H_0$ and $H_1$, respectively, are defined as follows:
\begin{equation*}\label{4.1}
  H_{0}:-T\leq\beta\leq T,
\end{equation*}
\begin{equation*}\label{4.2}
  H_{1}:\beta\notin[-T,T],
\end{equation*}
with the user-specified $[-T,T]$ interval, which we have taken without loss of generality to be symmetric with respect to zero, acting as ROPE.

The value of $T$ should in principle be chosen in such a way that ROPE contains values of $\beta$ that have small enough absolute magnitudes to be devoid of practical relevance (\citet{Kruschke2018}). Later in this paper we {\em calibrate} $T$ with respect to the data and to the model prior distribution. Moreover, use of an interval null hypothesis attenuates problems due to the model being (as all models) wrong, a consequence of this being sufficient amount of data will always cause the point null hypothesis to be rejected in favour of effects of negligible magnitude. The ROPE approach regards effects of small magnitude as practically ``equivalent to zero''.

\subsection{Causal discovery decision rule based on ROPE posterior odds}
\indent

Generally speaking, once the model has been specified, we would run one or more Hamiltonian MCMC chains (see \citet{Michael2017}) in the space of the unknown model parameters, and then focus
on the posterior samples for a particular (typically scalar) parameter
that informs us about the causal effect. One possibility is to consider the sampled values of parameter $\beta$ in Equations (\ref{2.3}-\ref{2.4}). Precisely, we would consider the proportion $V_1$ of sampled values of $\beta$ falling outside the ROPE, and the proportion $V_0$ of values falling inside the ROPE. The ratio of thecsevtwo quantities provides a
simulation consistent approximation to the posterior odds of
$\beta$ falling inside the ROPE, and an input to the following ternary {\em Causal Discovery Decision Rule} (CDDR) (\citet{Felix2018}):

\vspace{0.1cm}

\begin{itemize}

  \item If $\frac{V_{0}}{V_{1}} > 10$, accept the non-existence hypothesis with confidence;

  \item If $\frac{V_{0}}{V_{1}} < 0.1$, accept the existence hypothesis (and claim a causal discovery)
with confidence;

  \item If $0.1 \leq \frac{V_{0}}{V_{1}} \leq 10$, conclude in favour of an {\em uncertain evidence}
of a causal effect (uncertain outcome of the decision).

\end{itemize}

Now we discuss the prior distribution for the causal effect magnitude $\beta$. One possibility would be to assign $\beta$ a mixture prior distribution, that puts a $\pi_0$ probability on $\beta$ falling in the ROPE, and a complementary, $(1-\pi_0)$, probability on this parameter being drawn from a locally uninformative continuous distribution that we, without loss of generality, assume to be normal. Whenever we wish to express our prior ignorance about the existence of a causal effect, a reasonable value for $\pi_0$ is
$0.5$.
One way of
implementing the above prior within the model would be by expressing the prior distribution of $\beta$
conditional on an unobserved binary indicator which takes value $1$ in case of ``existence'' of the causal effect, and value $0$ in the case of ``non-existence'', so that this indicator would effectively act as a switch between the two components of the mixture prior. A causal discovery decision
rule could even be constructed on the basis of
such parameter, but we shall henceforth adopt the above CDDR
 based on the posterior samples of $\beta$. The samples are drawn
according to the procedure of the nect subsection.

\subsection{Importance sampling calculation of the posterior odds} \label{sec2.4}
\indent

The posterior samples of $\beta$ can be conveniently generated by using the following {\em mixture prior resampling} scheme. Let $\theta$ denote the full set of unknown quantities in the model (including parameters and missing data values) minus the causal effect $\beta$. The idea is to apply MCMC to a model where the ``true'' (mixture) prior for $\beta$, which we denote as $p^{true}(\beta)$, is replaced by a (computationally more convenient) continuous prior denoted as $p^{used}(\beta)$. A sample $S$ of values of $(\beta,\theta)$ will thus be generated from the ``incorrect'' posterior distribution
\begin{equation}\label{4.4}
        \begin{aligned}
  \pi^{used}(\beta,\theta \mid D) \;\; &\propto \;\;
p(D \mid \beta,\theta) \;\; p^{used}(\beta,\theta)
       \end{aligned}
\end{equation}
where the symbol $\propto$ stands for ``proportional to''
and $D$ denotes the data.

\vspace{0.1cm}

The idea is then to exploit principles of importance sampling in order to correct for the fact that we are sampling (\ref{4.4}) instead of the correct posterior. This is described in the following.

\vspace{0.1cm}

The \emph{true} posterior probability for $(\beta,\theta)$ is given, up to a proportionality constant we do not need to compute, by
\begin{equation*}\label{4.3}
  \pi^{true}(\beta,\theta \mid D)
\propto p(D \mid \beta,\theta) \; p^{true}(\beta,\theta).
\end{equation*}
By assuming $p^{true}(\beta,\theta) = p^{used}(\beta)p^{true}(\theta)$, the above equation can be re-written as
\begin{equation}\label{incorrect}
        \begin{aligned}
  \pi^{true}(\beta,\theta \mid D) &\propto
p(D \mid \beta,\theta) \;\; p^{used}(\beta,\theta) \;\;
\frac{p^{true}(\beta)}{p^{used}(\beta)} \\
 &\propto
 \pi^{used}(\beta,\theta \mid D) \;
\;\; \omega(\beta),
       \end{aligned}
\end{equation}
where $$\omega(\beta) \equiv\frac{p^{true}(\beta)}{p^{used}(\beta)}.$$

\vspace{0.1cm}

Let $S$ denote a set of $K$ samples of $(\beta,\theta)$, $$S\equiv\{\beta^{[k]},\theta^{[k]}\},~~~~~~k=1,\ldots,K,$$ generated from the ``incorrect'' posterior distribution $\pi^{used}(\beta,\theta \mid D)$. A reweighted resampling of $S$ with replacement, with the weight of each $k$th sample, $\omega^{[k]}$, obtained by evaluating $\omega(\beta)$ at it, $\omega^{[k]} \equiv \omega(\beta^{[k]})$, will yield a set of samples of $(\beta,\theta)$ we can think of as generated from the correct posterior. In particular, it will yield a set of posterior samples for $\beta$ which we can use
according to the CDDR of the preceding subsection to determine the test decision outcome. 

\vspace{0.1cm}

Recall our definition
of a prior for $\beta$ in
the form of a mixture of a locally uninformative normal distribution, say $N(0,10^{2})$, and a uniform density over $[-T,T]$. We call this mixture prior ``\emph{true} prior'' of $\beta$, denoted by $p^{true}(\beta)$. Consider an alternative model which only differs from the intended model by the fact that it replaces $p^{true}(\beta)$
with a computationally more manageable prior for $\beta$,
which we denote as $p^{used}(\beta)$, and, without loss of generality, take to be $N(0,10^{2})$. Then the choice $\pi_0=0.5$ leads to
\begin{equation*}\label{4.5}
  \omega^{[k]}=\frac{0.5\cdot N(\beta^{[k]}
\mid 0,10^{2})+0.5\cdot Unif(\beta^{[k]}
\mid -T,T)}{N(\beta^{[k]} \mid 0,10^{2})}.
\end{equation*}
where $Unif(q\mid -T,T)$ denotes the probability density at a real point $q$ under a rectangular (uniform) distribution with support $(-T, T)$. For a sample $\beta^{[k]}$ falling outside $[-T,T]$ this weight will be $0.5$. For a sample $\beta^{[k]}$ falling inside $[-T,T]$ it will be $0.5 + \frac{1}{4\cdot T\cdot N(\beta^{[k]} \mid 0,10^{2})}$. Hence samples falling outside the interval will be downweighted with respect to the samples inside, the downweighting being the more pronounced the smaller the interval.

\subsection{Calibration}\label{sec4}
\indent

\vspace{0.1cm}

Hypothesis test decision rules should be evaluated, more precisely {\em calibrated}, in accord with the principles of decision-theory, by using some measure of the {\em expected loss}. We shall consider {\em ternary} decisions with possible outcomes ``accept the hypothesis $H_1$ of existence of the causal effect'', ``accept the hypothesis $H_0$ of absence of the causal effect'' and ``uncertain outcome''. Let $L(\beta, A)$ denote the loss incurred when the true value of the causal parameter is $\beta$ and the chosen decision outcome is $A$. Let this function be defined as

\vspace{-0.3cm}

\begin{eqnarray*}
L(\beta=0, H_{0}
\;\mbox{accepted with confidence})\hspace{1cm}&=&0,\\
L(\beta=0, \mbox{uncertain outcome})\hspace{2.5cm} &=& a,\\
L(\beta=0, H_{1}
\;\mbox{accepted with confidence}
)\hspace{1cm}&=& 1,\\
L(\beta = \beta^{*}, H_{0}
\;\mbox{accepted with confidence}
)\hspace{0.8cm}&=&1,\\
L(\beta = \beta^{*}, \mbox{uncertain outcome})\hspace{2.3cm}&=& a,\\
L(\beta = \beta^{*}, H_{1}
\;\mbox{accepted with confidence}
)\hspace{0.8cm}&=& 0,
\end{eqnarray*}
where $\beta^{*} \neq 0$.

\vspace{0.1cm}

The choice of  $a$, with $0 \le a \le 1$, will depend on the applicative context. Large values of $a$ will be appropriate if conservative discoveries are desired at the cost of some decisions being held in a limbo.

\vspace{0.1cm}

The next section describes a simulation experiment where we compare results of a frequentist and of a Bayesian MR analysis of the same data,
by using different priors. In each of the simulated scenarios, the comparison is based on the expected loss:
\begin{eqnarray*}
L &\equiv &p(\mbox{uncertain outcome}
\mid \beta=0) \times
{\rm I}_{\beta=0}
\times a +\\
&&p(H_{1}
\;\mbox{accepted with confidence} \mid \beta=0)\times {\rm I}_{\beta=0}+ \\
&&p(H_{0}
\;\mbox{accepted with confidence} \mid \beta= \beta^{*}
) \times {\rm I}_{\beta=\beta^{*} }+ \\
&&p(\mbox{uncertain outcome} \mid \beta= \beta^{*}
) \; {\rm I}_{\beta=\beta^{*} } \;
\times a.
\end{eqnarray*}
with the probabilities
$p(.
\mid \beta=.)$
estimated by simulation. The first and last terms in the
expression of $L$ will be zero in the frequentist case. The intention in the experiment
will by no means be to prove that one of the two approaches is superior, but rather that use of a Bayesian approach with a ternary decision rule might often be a good idea, and that one
reason for this is that the frequentist method (but not the Bayesian one) is generally ``handicapped'' by its inability to allow for an uncertain outcome.

\vspace{0.1cm}

\section{Simulation experiment}
\indent

We are now going to describe a simulation experiment where
the expected loss criterion of the
preceding section is used to compare performances of a
frequentist and of a Bayesian MR method in realistic
data analysis scenarios. By allowing for
randomly missing values of the exposure variable,
we shall factor into the experiment the ability of the
Bayesian approach to coherently handle missing data, as
discussed elsewhere (\citet{Zou2020}).
No outcome-dependent selection mechanisms were simulated.

\vspace{0.2cm}

\subsection{Experiment design}\label{sec00}

We simulated situations where the following two datasets are jointly analysed:
\begin{itemize}

  \item Dataset $A$: collected from sample individuals with complete observations for $(\mathbf{Z}, X, Y)$;

  \item Dataset $B$: collected from individuals with completely observed values for $\mathbf{Z}$ and $Y$, and completely missing values of $X$.

\end{itemize}
We assumed no overlap, i.e., no individuals shared between $A$ and $B$.
\color{black}
Let the symbol $D_1$ denote the combined dataset $A \cup B$. In the special case where $B$ is empty, dataset $D_1$ lends itself to standard one-sample MR analysis. Analysis of $D_1$ will otherwise fall in the ``one-sample MR with missing
$X$-data'' category.
With reference to Equations (\ref{2.1})-(\ref{2.4}), we considered 18 different configurations:

\begin{itemize}

\item the rate of missingness of $X$: (80\%, 40\%, 0\%)

\item the strength of the $\balpha = (\alpha_{1}, \alpha_{2}, \ldots, \alpha_{J})$ coefficients, assumed to be the same for all IVs: $(\mathbf{0.3}$, $\mathbf{0.1}$, $\mathbf{0.05})$

\item the magnitude of the causal effect $\beta$: (0.3, 0)

\end{itemize}

In total, we considered
$3 \times 3 \times 2 = 18$ scenarios were simulated. Throughout the experiment, parameters $\delta_{X}$ and  $\delta_{Y}$ were set to 1 and the number $J$ of instruments was set to $15$. Two hundred datasets were simulated under each separate scenario, for a total of 3,600 datasets simulated during the experiment. Each of these 3,600 datasets was generated by the following sequence of steps:

\begin{enumerate}

\item simulate 1000 independent individuals characterized by realistic realizations of $\mathbf{Z}$ and then, on the basis of the $Z$s, generate for each individual values of $X$ and $Y$ in accord with Equations (\ref{2.1})-(\ref{2.4}). Call the resulting dataset $\textbf{H}$;

 \item randomly sample $n_A$ individuals from $\textbf{H}$, without replacement, and let the selected individuals, each with a completely observed ($\mathbf{Z}, X, Y$) vector, form the dataset that we have previously labelled as $A$;

 \item randomly sample $n_B$ individuals from $\textbf{H} \setminus A$ and take each of them to be characterized by observed ($\mathbf{Z}, Y$), with their corresponding values of $X$ treated as missing. Let these selected individuals form the dataset that we have previously labelled as $B$. At this point, we were ready to apply MR to data $D_1 = A \bigcup B$.

\end{enumerate}

The sample size of $D_1$ was set to be 400 throughout the experiment. Parameter $n_B$ was controlled by the rate of missingness of $X$ for the relevant scenario. For example, for a rate of missingness of $X$ of 80\%, we had $n_A = 80$ and $n_B = 320$. In the special case of a 0\% rate of missingness of $X$, it was $D_1 \equiv A$. MCMC computations were performed with the aid of the probabilistic programming language \texttt{Stan} (\citet{Stan2014, Martin2008}). Missing data imputation and causal effect estimation were performed simultaneously via MCMC, by exploiting the substantial equivalence of unknown model parameters and missing values in Bayesian analysis.

\vspace{0.2cm}

We compared our Bayesian method with the following two frequentist approaches to two-sample MR: two-stage least squares (2SLS) regression (\citet{Stephen2014}) in the case of a 0\% rate of missingness of $X$, and inverse-variance weighted (IVW) estimation (\citet{Jack2016}) otherwise. Application of frequentist IVW required the observed values of $Y$ in Dataset $A$ to be discarded to comply with the frequentist two-sample analysis mechanism. After each frequentist MR analysis of a simulated dataset, the null causal hypothesis $H_{0}:\beta=0$ was accepted iff the 95\% confidence interval for $\beta$ contained the null value 0. The alternative hypothesis $H_{1}:\beta \neq 0$ was otherwise accepted.

\vspace{0.2cm}

As far as the Bayesian analysis of each simulated dataset is concerned, we used the previously discussed computational procedure and
CDDR decision rule
to determine the outcome of the test, with the threshold $T$ and parameter $a$ randomly drawn from uniform distributions: $T \sim Unif(10^{-2}, 10^{-1})$ and $a \sim Unif(0, 0.6)$.
The information generated by the above procedure allowed
us to compute the expected losses for the MR methods under comparison, a high expected loss implying a high rate of false positives/negatives.

\subsection{Experiment results}\label{sec11}

\indent

\begin{figure}[h!]

\centering

  \includegraphics[scale=0.3]{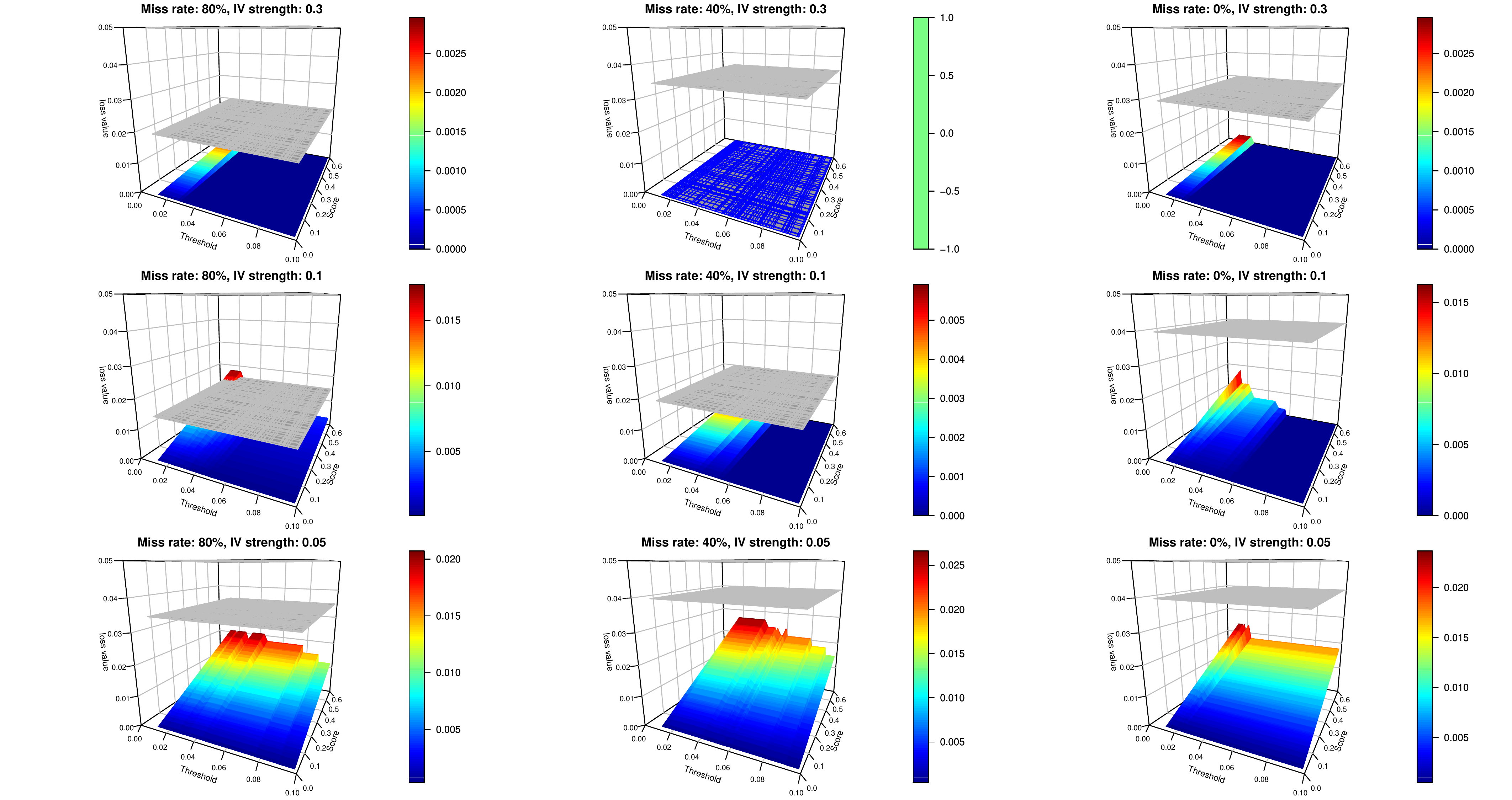}\\

  \caption{Loss of the frequentist and our Bayesian MR with binary $Y$ when $\beta = 0$. Each panel represent each combination of the missing rate of $X$ ($80\%,$ $40\%$ and $0\%$) and the IV strength (0.3, 0.1 and 0.05) based on 200 simulated datasets.}

  \label{figure71}

\end{figure}

Figure \ref{figure71} displays the results, when the $\beta = 0$, of the frequentist and Bayesian MR for each combination of IV strength and missing rate.  The flat grey surface in each panel depicts the loss of the frequentist MR and the coloured
surfaces
the loss of the Bayesian method. Unsurprisingly, the average loss incurred by the frequentist approach
does not appear depend on the values of $T$ or $a$, as these parameters are not involved in the calculation
of frequentist loss.

\begin{figure}[h!]

\centering

  \includegraphics[scale=0.3]{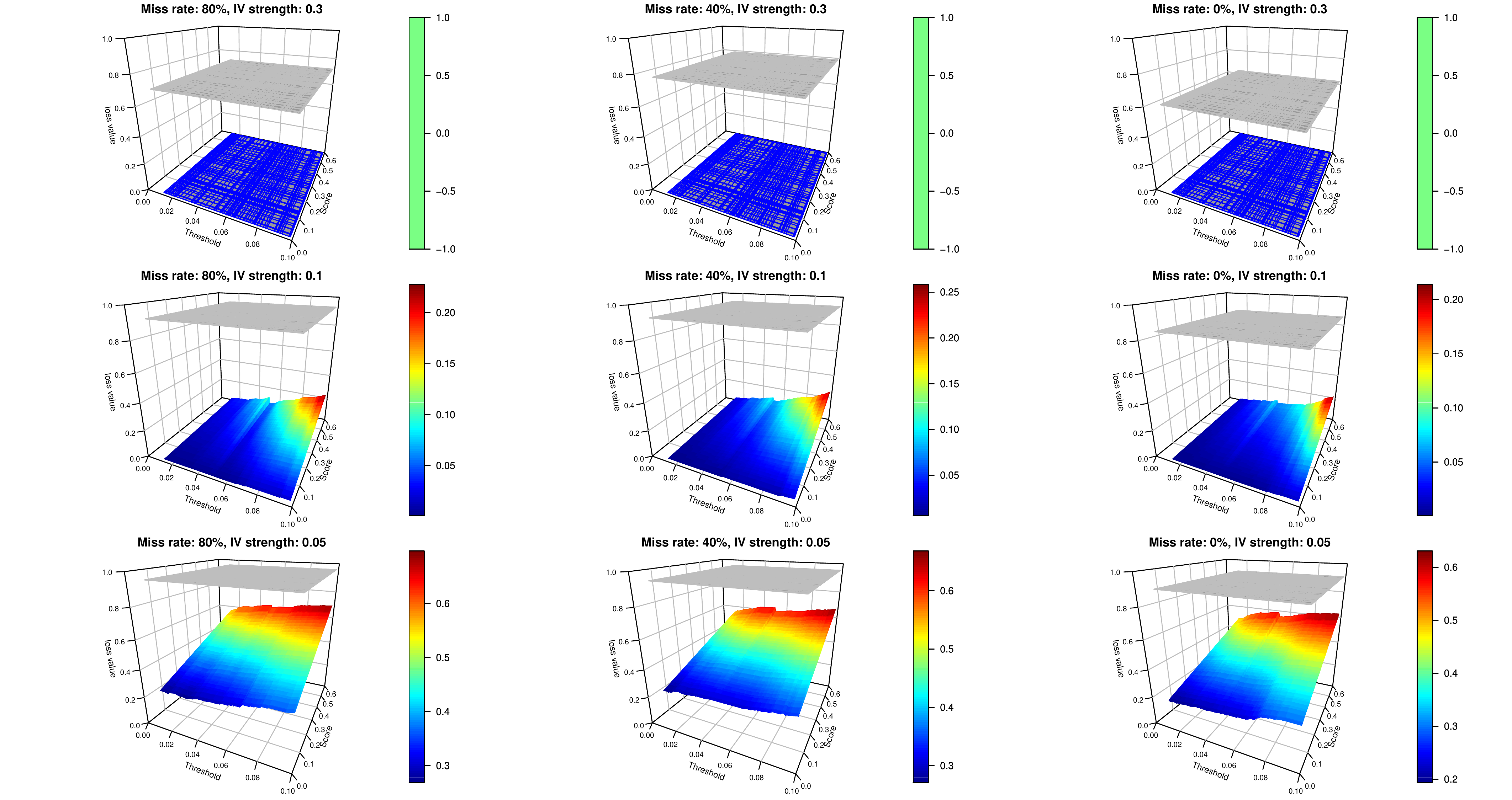}\\

  \caption{Loss of the frequentist (grey flat surfaces) and our Bayesian MR(coloured curves) with binary $Y$ when $\beta = 0.3$. Each panel represent each combination of the missing rate of $X$ ($80\%,$ $40\%$ and $0\%$) and the IV strength (0.3, 0.1 and 0.05) based on 200 simulated datasets.}

  \label{figure72}

\end{figure}

Our Bayesian method showed a lower loss almost uniformly across the configurations, with
the loss decreasing with an increasing
IV strength.
There were noticeable fluctuations of the loss when $T$ was small. This is because the weight $0.5 + \frac{1}{4\cdot T\cdot N(\beta^{[k]}\mid0,10^{2})}$ decreases as $T$ increases. When there were not enough samples falling in the wider tolerance interval to offset the decrease of weight, the loss became unstable. For example, considering two  intervals $[-T_{1},T_{1}]$ and $[-T_{2},T_{2}]$ ($T_{1} < T_{2}$), suppose there are 10 sampless falling in $[-T_{1},T_{1}]$, with weight $0.5 + \frac{1}{4\cdot T_{1}\cdot N(\beta^{[k]}\mid0,10^{2})}$, $k=1,\ldots,10$ (each point has a different density $N(\beta^{[k]}\mid0,10^{2})$). When $T$ increases from $T_{1}$ to $T_{2}$, the weights will reduce to $0.5 + \frac{1}{4\cdot T_{2}\cdot N(\beta^{[k]}\mid0,10^{2})}$ with the density values unchanged. Thus, if there are not enough new samples falling in $[-T_{2},T_{2}]$ to offset the decrease in weight, the loss will fluctuate.

\vspace{0.2cm}

When $\beta = 0.3$, our Bayesian method resulted in no loss when $\balpha = \mathbf{0.3}$ (Figure \ref{figure72}), showing a positive impact of higher IV strength on  decision-making. This is because no posterior samples fell in $[-T,T]$ for all different values of $T$. As IV strength decreased, the posterior distribution had a larger standard deviation and some samples fell in the interval for a large $T$, and we started to see a loss from the Bayesian method. When $T$ continued to increase, the wider interval  contained more samples, leading to a higher loss. When the level of IV strength decreased, the loss increased in our method. However, our method was still consistently better than the frequentist with a lower loss.

\vspace{0.2cm}

\section{Is Obesity a strong cause of Juvenile Myocardial Infarction?}\label{sec12}

\indent

With the improvement of living standards, recent decades have witnessed a dramatic increase in prevalence of obesity.  A common measure of obesity is the body mass index (BMI), defined as the weight in kilograms divided by the square of the height in meters.
A high value of BMI is taken to indicate an excess of body fat.
Studies have demonstrated that obesity acts as a major causal risk
factor for cardiovascular disease and hypertension. To
the best of our knowledge, no studies have examined these links by
an analysis that focuses on occurrence of MI {\em at an early age}.
Yet, early age is where
the influence
of genetics on cardiovascular events is at its highest, which
makes a MR analysis of
the causes of early-age MI most appealing and less vulnerable to biases.
Motivated by these considerations, our study focuses on the causal effect of BMI on
occurrence of a MI before age 45.

\vspace{0.2cm}

Our analysis was based on data from an
Italian study of the genetics of infarction (\citet{Carlo2012}).
JMI cases were ascertained on the basis of hospitalization
for acute myocardial infarction between ages 40 and 45,
from 1996 to 2002. The recorded values of BMI, measured after occurrence of
JMI, were considered representative of pre-JMI obesity level.

\vspace{0.2cm}

One caveat in the analysis we are going to describe is that post-JMI
BMI levels may reflect recent changes in the patient's
lifestyle and behaviour, resulting in rather weak genetic associations
with BMI and, as a consequence, in higher vulnerability to bias.

\vspace{0.2cm}

\begin{figure}[h!]

\centering

\includegraphics[scale=0.3]{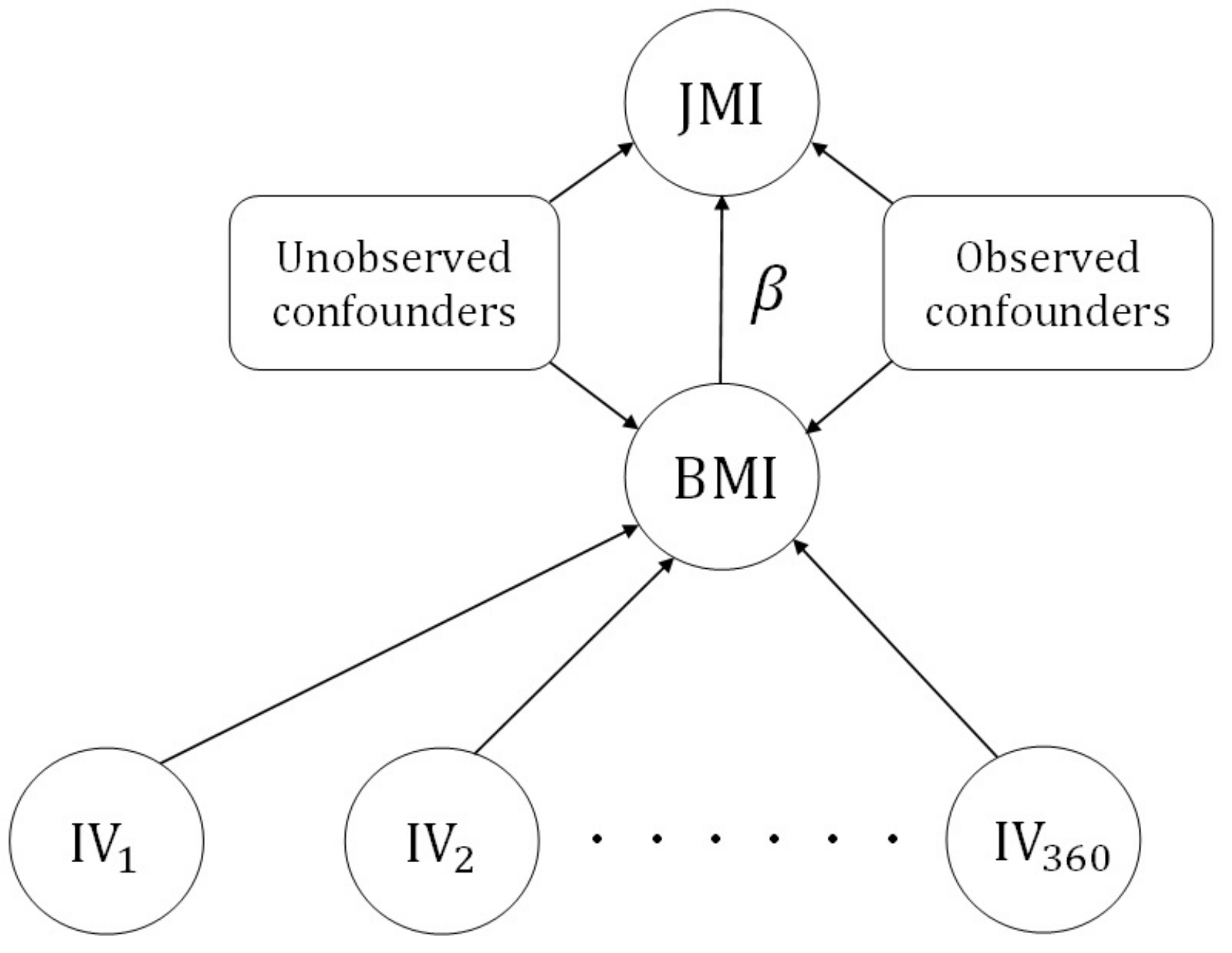}\\

\caption{MR model with 360 instrumental variables, ($\rm IV_{1}$, $\rm IV_{2}$, \ldots, $\rm IV_{360}$), used in our illustrative study to assess the causal effect of BMI on juvenile myocardial infarction (JMI). Observed values of 5 potential confounders (sex, smoking status, alcohol consumption, cocaine consumption and age), were included in the model (see main text). The model assumes there are no associations between IVs and JMI other than those mediated by BMI.}

\label{figure4}

\end{figure}

A graphical representation of the adopted MR model
is shown in Figure \ref{figure4}. The narrow age
range of the event in our study attenuates problems
introduced by censoring, and justifies our choice of
representing disease outcome as a binary
variable (1: had MI, 0: did not have MI) in our analysis.

\vspace{0.2cm}

The single nucleotide polymorphisms (SNPs) associated with BMI were identified based on the datasets from genome-wide association study (GWAS) and UK Biobank\footnote{We use the datasets from https://www.ebi.ac.uk/gwas/publications/25673413 and http://biobank.ctsu.ox.ac.uk/crystal/field.cgi?id=21001.} ($P\leq 5\times10^{-8}$). Through the command-line program \texttt{Plink} (\citet{PLINK}), as many as 360 independent ($r^{2} < 0.001$) SNPs were selected and then used as instruments for assessment of obesity causal effect on JMI. SNPs were coded as 3 valued $(0,1,2)$ counts of the minor allele, after appropriate cross-study harmonization. In total, 521 independent individuals were studied in the illustrative analysis based on our Bayesian method.
Values of BMI were standardised (mean 0, standard deviation 1) prior to analysis.
The observed values of the following 5 potential confounders were  included
in the model: sex (Male/Female), smoking status (Yes/No) alcohol
consumption (Yes/No), cocaine consumption (Yes/No) and age.

\begin{center}
\begin{figure}
\centering
\subfigure[Posterior distribution curve for $\beta$.]{
\centering
\begin{minipage}[b]{10cm}
  \includegraphics[scale=0.15]{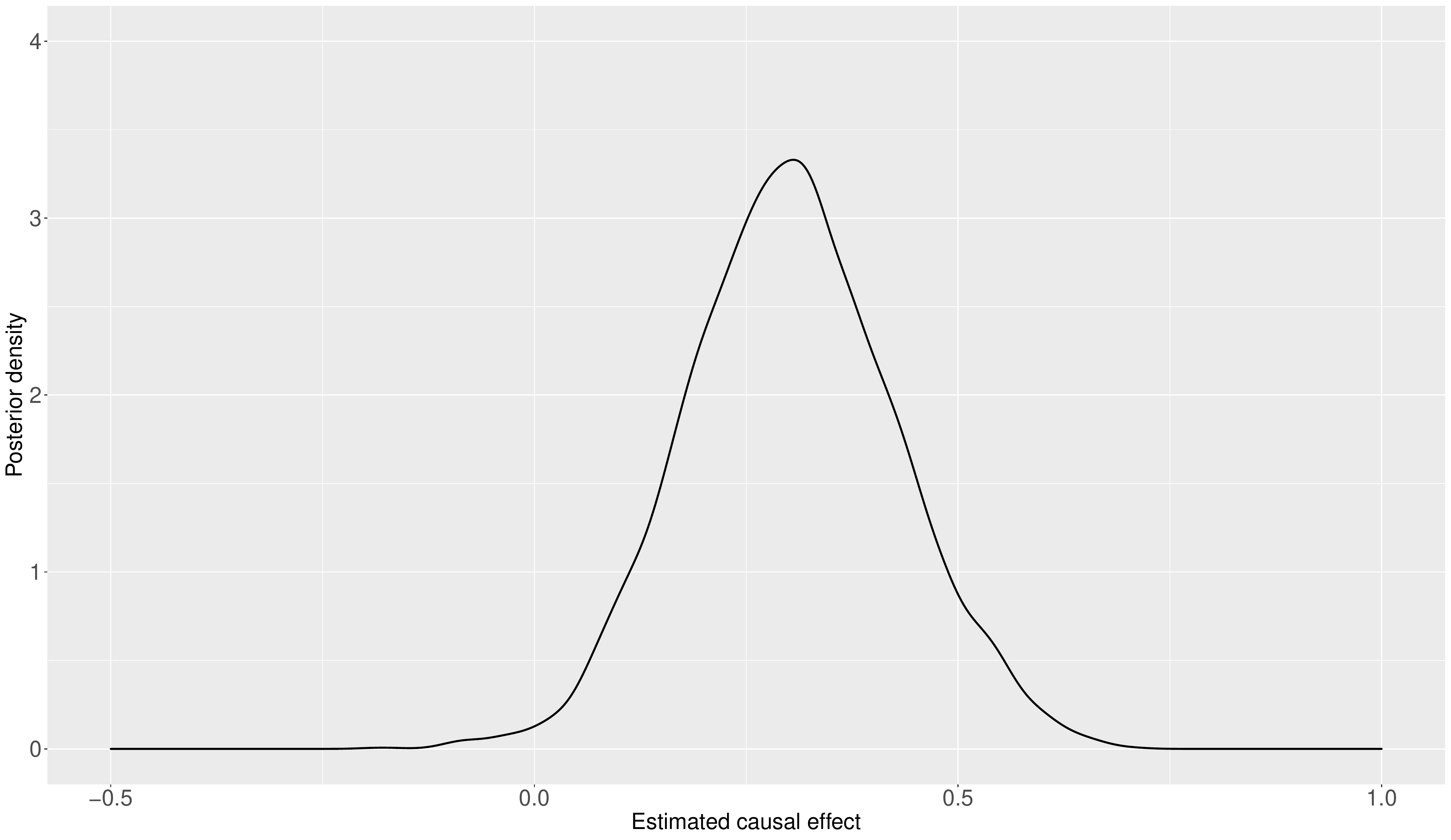}
\end{minipage}
}
\\
\subfigure[Bayesian posterior parameter trace plot for $\beta$.]{
\centering
\begin{minipage}[b]{10cm}
  \includegraphics[scale=0.15]{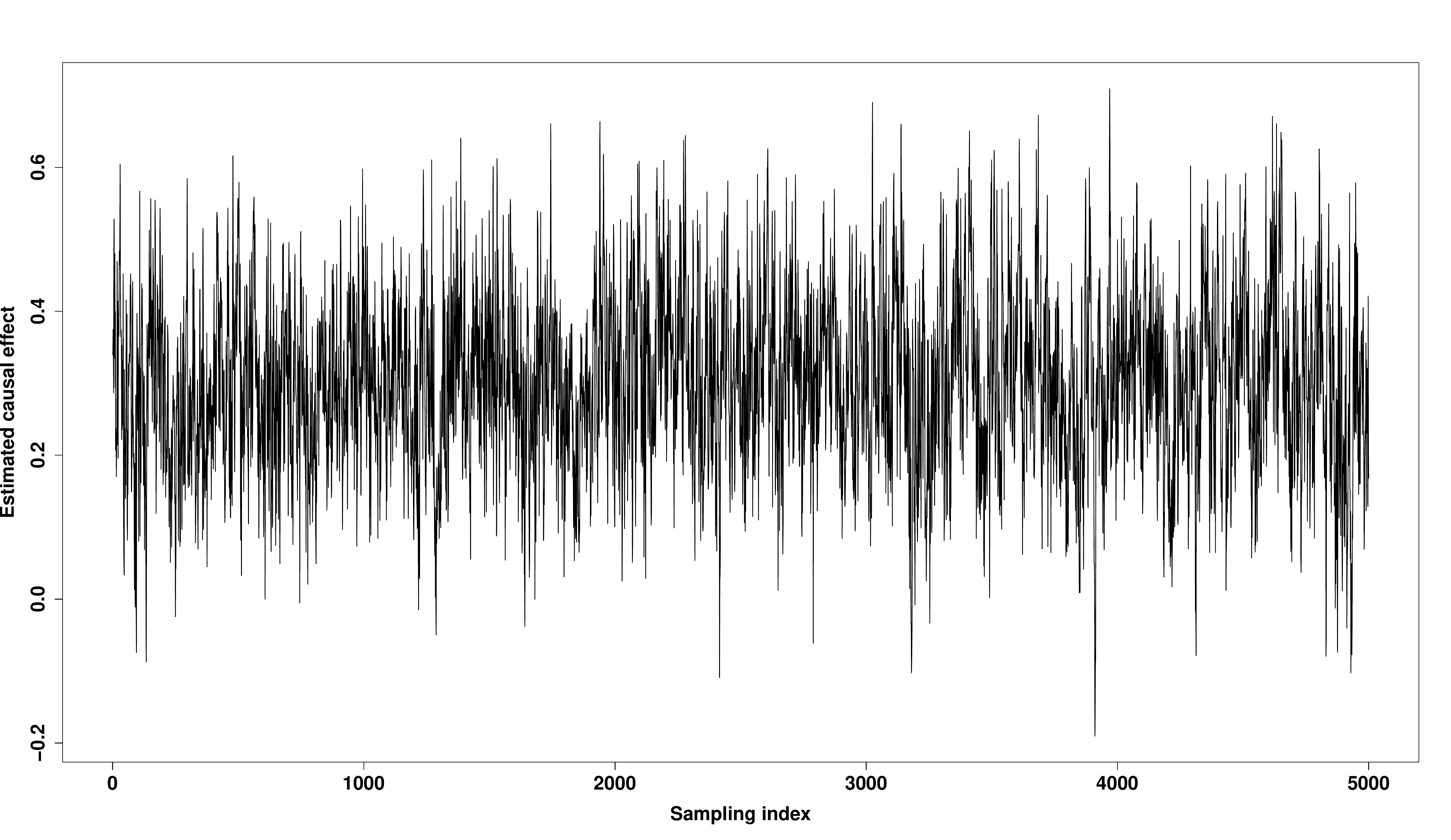}
\end{minipage}
}
\caption{\small Estimated causal effect of {\em standardized} BMI on JMI, based on the Bayesian Mendelian randomization analysis we have performed data from our illustrative study. $(a)$ Posterior distribution curve for causal parameter $\beta$ with posterior mean 0.303 and 95\% credible interval (0.069, 0.550). $(b)$ Bayesian posterior parameter trace plot for $\beta$.} 
\label{figure66}
\end{figure}
\end{center}

We ran the Markov chain in the space of the model
unknowns (model parameters and missing BMI values).
The chain was 20,000 iterations long, the last 5,000 iterations being used for purposes of inference. Figure \ref{figure66} shows superimposed posterior density plot (Panel (a)) and Bayesian posterior parameter trace plot (Panel (b)) for the causal effect of standardized BMI on JMI. The trace in (b) indicates reasonably good mixing of the chain, with $\widehat{R}=1.002$. For each
of the thresholds $(0.02, 0.04, 0.06, 0.08)$ for $T$, after
the  {\em mixture prior resampling} discussed in Section \ref{sec2.4},
we had $0.1 < \frac{V_{0}}{V_{1}} < 10$. In the light of this, and in spite
of the 95\% credible interval for $\beta$ lying entirely in the positive real axis,
our test decision was {\em uncertain evidence} in favour of a causal
effect of genetically induced changes in obesity on JMI.

\section{Discussion}
\indent

In our approach to Bayesian MR analysis,
the hypothesis $H_0$ of {\em non-existence} of the causal effect
of interest is represented by a user-specified
interval of values of the causal effect, which we may refer to
by using the established term "region of practical equivalence" (ROPE). Importance sampling
technology is used to approximate the posterior
odds of the causal effect falling inside this interval. A sufficiently
large value of this posterior odds will lead to
acceptance of the null no-effect hypothesis, whereas a sufficiently small
value will lead to acceptance of the alternative hypothesis, and to a causal discovery claim.
A third, ``uncertain'', decision outcome is available for situations where
the posterior odds is neither large nor small
enough, indicating scarce data support to either hypothesis. The uncertain outcome has been introduced to reduce chances of placing
undue confidence in a hypothesis that is only weakly supported from the data
 \color{black}. We have incorporated this ternary test decision logic into the
Bayesian MR framework proposed by \citet{Carlo2018} and further refined by \citet{Zou2020}.
The decision rule can be calibrated via simulation  \color{black} by acting
on the differential weighting parameters of a loss function.

\vspace{0.2cm}

In a simulation experiment, we have compared our method with a standard MR method in terms of expected loss, by allowing loss function parameters to vary within reasonably wide intervals. The experiment suggests that, within the examined scenarios,
our method outperforms standard MR, and this may be due to
 the latter being handicapped by inability to accommodate decision
uncertainty. We consider  our proposed method as a contribution to
research on more reproducible MR analysis.

\vspace{0.2cm}

We have applied our proposed method to a MR study of the causal effect
of obesity on {\em juvenile} myocardial infarction, based on a unique
dataset. The study concludes in favour of
an {\em uncertain evidence} of a non-null causal effect.

\vspace{0.5cm}

\comm{

\section{Results}\label{sec2}

Sample body text. Sample body text. Sample body text. Sample body text. Sample body text. Sample body text. Sample body text. Sample body text.

\section{This is an example for first level head---section head}\label{sec3}

\subsection{This is an example for second level head---subsection head}\label{subsec2}

\subsubsection{This is an example for third level head---subsubsection head}\label{subsubsec2}

Sample body text. Sample body text. Sample body text. Sample body text. Sample body text. Sample body text. Sample body text. Sample body text.

\section{Equations}\label{sec4}

\begin{align}
 X \mid \mathbf{Z},U &\sim N\left(\sum\limits_{k=1}^{J}\alpha_{k}Z_{k}+\delta_{X}U,\sigma_{X}^{2}\right)
\end{align}

Equations in \LaTeX\ can either be inline or on-a-line by itself (``display equations''). For
inline equations use the \verb+$...$+ commands. E.g.: The equation
$H\psi = E \psi$ is written via the command \verb+$H \psi = E \psi$+.

For display equations (with auto generated equation numbers)
one can use the equation or align environments:
\begin{equation}
\|\tilde{X}(k)\|^2 \leq\frac{\sum\limits_{i=1}^{p}\left\|\tilde{Y}_i(k)\right\|^2+\sum\limits_{j=1}^{q}\left\|\tilde{Z}_j(k)\right\|^2 }{p+q}.\label{eq1}
\end{equation}
where,
\begin{align}
\mathbf{D}_\mu &=  \partial_\mu - ig \frac{\lambda^a}{2} A^a_\mu \nonumber \\
F^a_{\mu\nu} &= \partial_\mu A^a_\nu - \partial_\nu A^a_\mu + g f^{abc} A^b_\mu A^a_\nu \label{eq2}
\end{align}
Notice the use of \verb+\nonumber+ in the align environment at the end
of each line, except the last, so as not to produce equation numbers on
lines where no equation numbers are required. The \verb+\label{}+ command
should only be used at the last line of an align environment where
\verb+\nonumber+ is not used.
\begin{equation}
Y_\infty = \left( \frac{m}{\textrm{GeV}} \right)^{-3}
    \left[ 1 + \frac{3 \ln(m/\textrm{GeV})}{15}
    + \frac{\ln(c_2/5)}{15} \right]
\end{equation}
The class file also supports the use of \verb+\mathbb{}+, \verb+\mathscr{}+ and
\verb+\mathcal{}+ commands. As such \verb+\mathbb{R}+, \verb+\mathscr{R}+
and \verb+\mathcal{R}+ produces $\mathbb{R}$, $\mathscr{R}$ and $\mathcal{R}$
respectively (refer Subsubsection~\ref{subsubsec2}).

\section{Tables}\label{sec5}

Tables can be inserted via the normal table and tabular environment. To put
footnotes inside tables you should use \verb+\footnotetext[]{...}+ tag.
The footnote appears just below the table itself (refer Tables~\ref{tab1} and \ref{tab2}).
For the corresponding footnotemark use \verb+\footnotemark[...]+

\begin{table}[h]
\begin{center}
\begin{minipage}{174pt}
\caption{Caption text}\label{tab1}%
\begin{tabular}{@{}llll@{}}
\toprule
Column 1 & Column 2  & Column 3 & Column 4\\
\midrule
row 1    & data 1   & data 2  & data 3  \\
row 2    & data 4   & data 5\footnotemark[1]  & data 6  \\
row 3    & data 7   & data 8  & data 9\footnotemark[2]  \\
\botrule
\end{tabular}
\footnotetext{Source: This is an example of table footnote. This is an example of table footnote.}
\footnotetext[1]{Example for a first table footnote. This is an example of table footnote.}
\footnotetext[2]{Example for a second table footnote. This is an example of table footnote.}
\end{minipage}
\end{center}
\end{table}

\noindent
The input format for the above table is as follows:

\bigskip
\begin{verbatim}
\begin{table}[<placement-specifier>]
\begin{center}
\begin{minipage}{<preferred-table-width>}
\caption{<table-caption>}\label{<table-label>}%
\begin{tabular}{@{}llll@{}}
\toprule
Column 1 & Column 2 & Column 3 & Column 4\\
\midrule
row 1 & data 1 & data 2 & data 3 \\
row 2 & data 4 & data 5\footnotemark[1] & data 6 \\
row 3 & data 7 & data 8 & data 9\footnotemark[2]\\
\botrule
\end{tabular}
\footnotetext{Source: This is an example of table footnote.
This is an example of table footnote.}
\footnotetext[1]{Example for a first table footnote.
This is an example of table footnote.}
\footnotetext[2]{Example for a second table footnote.
This is an example of table footnote.}
\end{minipage}
\end{center}
\end{table}
\end{verbatim}
\bigskip

\begin{table}[h]
\begin{center}
\begin{minipage}{\textwidth}
\caption{Example of a lengthy table which is set to full textwidth}\label{tab2}
\begin{tabular*}{\textwidth}{@{\extracolsep{\fill}}lcccccc@{\extracolsep{\fill}}}
\toprule%
& \multicolumn{3}{@{}c@{}}{Element 1\footnotemark[1]} & \multicolumn{3}{@{}c@{}}{Element 2\footnotemark[2]} \\\cmidrule{2-4}\cmidrule{5-7}%
Project & Energy & $\sigma_{calc}$ & $\sigma_{expt}$ & Energy & $\sigma_{calc}$ & $\sigma_{expt}$ \\
\midrule
Element 3  & 990 A & 1168 & $1547\pm12$ & 780 A & 1166 & $1239\pm100$\\
Element 4  & 500 A & 961  & $922\pm10$  & 900 A & 1268 & $1092\pm40$\\
\botrule
\end{tabular*}
\footnotetext{Note: This is an example of table footnote. This is an example of table footnote this is an example of table footnote this is an example of~table footnote this is an example of table footnote.}
\footnotetext[1]{Example for a first table footnote.}
\footnotetext[2]{Example for a second table footnote.}
\end{minipage}
\end{center}
\end{table}

In case of double column layout, tables which do not fit in single column width should be set to full text width. For this, you need to use \verb+\begin{table*}+ \verb+...+ \verb+\end{table*}+ instead of \verb+\begin{table}+ \verb+...+ \verb+\end{table}+ environment. Lengthy tables which do not fit in textwidth should be set as rotated table. For this, you need to use \verb+\begin{sidewaystable}+ \verb+...+ \verb+\end{sidewaystable}+ instead of \verb+\begin{table*}+ \verb+...+ \verb+\end{table*}+ environment. This environment puts tables rotated to single column width. For tables rotated to double column width, use \verb+\begin{sidewaystable*}+ \verb+...+ \verb+\end{sidewaystable*}+.

\begin{sidewaystable}
\sidewaystablefn%
\begin{center}
\begin{minipage}{\textheight}
\caption{Tables which are too long to fit, should be written using the ``sidewaystable'' environment as shown here}\label{tab3}
\begin{tabular*}{\textheight}{@{\extracolsep{\fill}}lcccccc@{\extracolsep{\fill}}}
\toprule%
& \multicolumn{3}{@{}c@{}}{Element 1\footnotemark[1]}& \multicolumn{3}{@{}c@{}}{Element\footnotemark[2]} \\\cmidrule{2-4}\cmidrule{5-7}%
Projectile & Energy & $\sigma_{calc}$ & $\sigma_{expt}$ & Energy & $\sigma_{calc}$ & $\sigma_{expt}$ \\
\midrule
Element 3 & 990 A & 1168 & $1547\pm12$ & 780 A & 1166 & $1239\pm100$ \\
Element 4 & 500 A & 961  & $922\pm10$  & 900 A & 1268 & $1092\pm40$ \\
Element 5 & 990 A & 1168 & $1547\pm12$ & 780 A & 1166 & $1239\pm100$ \\
Element 6 & 500 A & 961  & $922\pm10$  & 900 A & 1268 & $1092\pm40$ \\
\botrule
\end{tabular*}
\footnotetext{Note: This is an example of table footnote this is an example of table footnote this is an example of table footnote this is an example of~table footnote this is an example of table footnote.}
\footnotetext[1]{This is an example of table footnote.}
\end{minipage}
\end{center}
\end{sidewaystable}

\section{Figures}\label{sec6}

As per the \LaTeX\ standards you need to use eps images for \LaTeX\ compilation and \verb+pdf/jpg/png+ images for \verb+PDFLaTeX+ compilation. This is one of the major difference between \LaTeX\ and \verb+PDFLaTeX+. Each image should be from a single input .eps/vector image file. Avoid using subfigures. The command for inserting images for \LaTeX\ and \verb+PDFLaTeX+ can be generalized. The package used to insert images in \verb+LaTeX/PDFLaTeX+ is the graphicx package. Figures can be inserted via the normal figure environment as shown in the below example:

\bigskip
\begin{verbatim}
\begin{figure}[<placement-specifier>]
\centering
\includegraphics{<eps-file>}
\caption{<figure-caption>}\label{<figure-label>}
\end{figure}
\end{verbatim}
\bigskip

\begin{figure}[h]%
\centering
\includegraphics[width=0.9\textwidth]{plot4.pdf}
\caption{This is a widefig. This is an example of long caption this is an example of long caption  this is an example of long caption this is an example of long caption}\label{fig1}
\end{figure}

In case of double column layout, the above format puts figure captions/images to single column width. To get spanned images, we need to provide \verb+\begin{figure*}+ \verb+...+ \verb+\end{figure*}+.

For sample purpose, we have included the width of images in the optional argument of \verb+\includegraphics+ tag. Please ignore this.

\section{Algorithms, Program codes and Listings}\label{sec7}

Packages \verb+algorithm+, \verb+algorithmicx+ and \verb+algpseudocode+ are used for setting algorithms in \LaTeX\ using the format:

\bigskip
\begin{verbatim}
\begin{algorithm}
\caption{<alg-caption>}\label{<alg-label>}
\begin{algorithmic}[1]
. . .
\end{algorithmic}
\end{algorithm}
\end{verbatim}
\bigskip

You may refer above listed package documentations for more details before setting \verb+algorithm+ environment. For program codes, the ``program'' package is required and the command to be used is \verb+\begin{program}+ \verb+...+ \verb+\end{program}+. A fast exponentiation procedure:

\begin{program}
\BEGIN \\ %
  \FOR i:=1 \TO 10 \STEP 1 \DO
     |expt|(2,i); \\ |newline|() \OD %
\rcomment{Comments will be set flush to the right margin}
\WHERE
\PROC |expt|(x,n) \BODY
          z:=1;
          \DO \IF n=0 \THEN \EXIT \FI;
             \DO \IF |odd|(n) \THEN \EXIT \FI;
\COMMENT{This is a comment statement};
                n:=n/2; x:=x*x \OD;
             \{ n>0 \};
             n:=n-1; z:=z*x \OD;
          |print|(z) \ENDPROC
\END
\end{program}

\begin{algorithm}
\caption{Calculate $y = x^n$}\label{algo1}
\begin{algorithmic}[1]
\Require $n \geq 0 \vee x \neq 0$
\Ensure $y = x^n$
\State $y \Leftarrow 1$
\If{$n < 0$}\label{algln2}
        \State $X \Leftarrow 1 / x$
        \State $N \Leftarrow -n$
\Else
        \State $X \Leftarrow x$
        \State $N \Leftarrow n$
\EndIf
\While{$N \neq 0$}
        \If{$N$ is even}
            \State $X \Leftarrow X \times X$
            \State $N \Leftarrow N / 2$
        \Else[$N$ is odd]
            \State $y \Leftarrow y \times X$
            \State $N \Leftarrow N - 1$
        \EndIf
\EndWhile
\end{algorithmic}
\end{algorithm}
\bigskip

Similarly, for \verb+listings+, use the \verb+listings+ package. \verb+\begin{lstlisting}+ \verb+...+ \verb+\end{lstlisting}+ is used to set environments similar to \verb+verbatim+ environment. Refer to the \verb+lstlisting+ package documentation for more details.

\bigskip
\begin{minipage}{\hsize}%
\lstset{frame=single,framexleftmargin=-1pt,framexrightmargin=-17pt,framesep=12pt,linewidth=0.98\textwidth,language=pascal}
\begin{lstlisting}
for i:=maxint to 0 do
begin
{ do nothing }
end;
Write('Case insensitive ');
Write('Pascal keywords.');
\end{lstlisting}
\end{minipage}

\section{Cross referencing}\label{sec8}

Environments such as figure, table, equation and align can have a label
declared via the \verb+\label{#label}+ command. For figures and table
environments use the \verb+\label{}+ command inside or just
below the \verb+\caption{}+ command. You can then use the
\verb+\ref{#label}+ command to cross-reference them. As an example, consider
the label declared for Figure~\ref{fig1} which is
\verb+\label{fig1}+. To cross-reference it, use the command
\verb+Figure \ref{fig1}+, for which it comes up as
``Figure~\ref{fig1}''.

To reference line numbers in an algorithm, consider the label declared for the line number 2 of Algorithm~\ref{algo1} is \verb+\label{algln2}+. To cross-reference it, use the command \verb+\ref{algln2}+ for which it comes up as line~\ref{algln2} of Algorithm~\ref{algo1}.

\subsection{Details on reference citations}\label{subsec7}

Standard \LaTeX\ permits only numerical citations. To support both numerical and author-year citations this template uses \verb+natbib+ \LaTeX\ package. For style guidance please refer to the template user manual.

Here is an example for \verb+\cite{...}+: \cite{bib1}. Another example for \verb+\citep{...}+: \citep{bib2}. For author-year citation mode, \verb+\cite{...}+ prints Jones et al. (1990) and \verb+\citep{...}+ prints (Jones et al., 1990).

All cited bib entries are printed at the end of this article: \cite{bib3}, \cite{bib4}, \cite{bib5}, \cite{bib6}, \cite{bib7}, \cite{bib8}, \cite{bib9}, \cite{bib10}, \cite{bib11} and \cite{bib12}.

\section{Examples for theorem like environments}\label{sec10}

For theorem like environments, we require \verb+amsthm+ package. There are three types of predefined theorem styles exists---\verb+thmstyleone+, \verb+thmstyletwo+ and \verb+thmstylethree+

\bigskip
\begin{tabular}{|l|p{19pc}|}
\hline
\verb+thmstyleone+ & Numbered, theorem head in bold font and theorem text in italic style \\\hline
\verb+thmstyletwo+ & Numbered, theorem head in roman font and theorem text in italic style \\\hline
\verb+thmstylethree+ & Numbered, theorem head in bold font and theorem text in roman style \\\hline
\end{tabular}
\bigskip

For mathematics journals, theorem styles can be included as shown in the following examples:

\begin{theorem}[Theorem subhead]\label{thm1}
Example theorem text. Example theorem text. Example theorem text. Example theorem text. Example theorem text.
Example theorem text. Example theorem text. Example theorem text. Example theorem text. Example theorem text.
Example theorem text.
\end{theorem}

Sample body text. Sample body text. Sample body text. Sample body text. Sample body text. Sample body text. Sample body text. Sample body text.

\begin{proposition}
Example proposition text. Example proposition text. Example proposition text. Example proposition text. Example proposition text.
Example proposition text. Example proposition text. Example proposition text. Example proposition text. Example proposition text.
\end{proposition}

Sample body text. Sample body text. Sample body text. Sample body text. Sample body text. Sample body text. Sample body text. Sample body text.

\begin{example}
Phasellus adipiscing semper elit. Proin fermentum massa
ac quam. Sed diam turpis, molestie vitae, placerat a, molestie nec, leo. Maecenas lacinia. Nam ipsum ligula, eleifend
at, accumsan nec, suscipit a, ipsum. Morbi blandit ligula feugiat magna. Nunc eleifend consequat lorem.
\end{example}

Sample body text. Sample body text. Sample body text. Sample body text. Sample body text. Sample body text. Sample body text. Sample body text.

\begin{remark}
Phasellus adipiscing semper elit. Proin fermentum massa
ac quam. Sed diam turpis, molestie vitae, placerat a, molestie nec, leo. Maecenas lacinia. Nam ipsum ligula, eleifend
at, accumsan nec, suscipit a, ipsum. Morbi blandit ligula feugiat magna. Nunc eleifend consequat lorem.
\end{remark}

Sample body text. Sample body text. Sample body text. Sample body text. Sample body text. Sample body text. Sample body text. Sample body text.

\begin{definition}[Definition sub head]
Example definition text. Example definition text. Example definition text. Example definition text. Example definition text. Example definition text. Example definition text. Example definition text.
\end{definition}

Additionally a predefined ``proof'' environment is available: \verb+\begin{proof}+ \verb+...+ \verb+\end{proof}+. This prints a ``Proof'' head in italic font style and the ``body text'' in roman font style with an open square at the end of each proof environment.

\begin{proof}
Example for proof text. Example for proof text. Example for proof text. Example for proof text. Example for proof text. Example for proof text. Example for proof text. Example for proof text. Example for proof text. Example for proof text.
\end{proof}

Sample body text. Sample body text. Sample body text. Sample body text. Sample body text. Sample body text. Sample body text. Sample body text.

\begin{proof}[Proof of Theorem~{\upshape\ref{thm1}}]
Example for proof text. Example for proof text. Example for proof text. Example for proof text. Example for proof text. Example for proof text. Example for proof text. Example for proof text. Example for proof text. Example for proof text.
\end{proof}

\noindent
For a quote environment, use \verb+\begin{quote}...\end{quote}+
\begin{quote}
Quoted text example. Aliquam porttitor quam a lacus. Praesent vel arcu ut tortor cursus volutpat. In vitae pede quis diam bibendum placerat. Fusce elementum
convallis neque. Sed dolor orci, scelerisque ac, dapibus nec, ultricies ut, mi. Duis nec dui quis leo sagittis commodo.
\end{quote}

Sample body text. Sample body text. Sample body text. Sample body text. Sample body text (refer Figure~\ref{fig1}). Sample body text. Sample body text. Sample body text (refer Table~\ref{tab3}).

\section{Methods}\label{sec11}

Topical subheadings are allowed. Authors must ensure that their Methods section includes adequate experimental and characterization data necessary for others in the field to reproduce their work. Authors are encouraged to include RIIDs where appropriate.

\textbf{Ethical approval declarations} (only required where applicable) Any article reporting experiment/s carried out on (i)~live vertebrate (or higher invertebrates), (ii)~humans or (iii)~human samples must include an unambiguous statement within the methods section that meets the following requirements:

\begin{enumerate}[1.]
\item Approval: a statement which confirms that all experimental protocols were approved by a named institutional and/or licensing committee. Please identify the approving body in the methods section

\item Accordance: a statement explicitly saying that the methods were carried out in accordance with the relevant guidelines and regulations

\item Informed consent (for experiments involving humans or human tissue samples): include a statement confirming that informed consent was obtained from all participants and/or their legal guardian/s
\end{enumerate}

If your manuscript includes potentially identifying patient/participant information, or if it describes human transplantation research, or if it reports results of a clinical trial then  additional information will be required. Please visit (\url{https://www.nature.com/nature-research/editorial-policies}) for Nature Portfolio journals, (\url{https://www.springer.com/gp/authors-editors/journal-author/journal-author-helpdesk/publishing-ethics/14214}) for Springer Nature journals, or (\url{https://www.biomedcentral.com/getpublished/editorial-policies\#ethics+and+consent}) for BMC.

\section{Discussion}\label{sec12}

Discussions should be brief and focused. In some disciplines use of Discussion or `Conclusion' is interchangeable. It is not mandatory to use both. Some journals prefer a section `Results and Discussion' followed by a section `Conclusion'. Please refer to Journal-level guidance for any specific requirements.

\section{Conclusion}\label{sec13}

Conclusions may be used to restate your hypothesis or research question, restate your major findings, explain the relevance and the added value of your work, highlight any limitations of your study, describe future directions for research and recommendations.

In some disciplines use of Discussion or 'Conclusion' is interchangeable. It is not mandatory to use both. Please refer to Journal-level guidance for any specific requirements.

}

\backmatter

\comm{

\bmhead{Supplementary information}

If your article has accompanying supplementary file/s please state so here.

Authors reporting data from electrophoretic gels and blots should supply the full unprocessed scans for key as part of their Supplementary information. This may be requested by the editorial team/s if it is missing.

Please refer to Journal-level guidance for any specific requirements.

}

\bmhead{Acknowledgments}

We thank Philip Dawid for advice on methodological aspects of the work, and Luisa Bernardinelli and Diego Ardissino for providing the data for the Illustrative Study and contributing to the interpretation of the results. Any misinterpretation is, of course, entirely a responsibility of the authors.

\comm{

We thank Philip Dawid for advice on methodological aspects of the work and Luisa Bernardinelli for suggesting and providing the necessary support in the conception and data collection stages of the applicative study. Any misinterpretation is, of course, entirely a responsibility of the authors.

}

\vspace{0.2cm}

\section*{Declarations}

\comm{

Some journals require declarations to be submitted in a standardised format. Please check the Instructions for Authors of the journal to which you are submitting to see if you need to complete this section. If yes, your manuscript must contain the following sections under the heading `Declarations':

}

\textbf{Funding} This work was funded by Manchester-CSC. The funder had no role in study design, data generation and statistical analysis, interpretation of data or preparation of the manuscript.

\noindent
\textbf{Competing interests} The authors declare that they have no competing interests.

\noindent
\textbf{Ethics approval} Not applicable

\noindent
\textbf{Consent to participate} Not applicable

\noindent
\textbf{Consent for publication} Not applicable

\noindent
\textbf{Availability of data and materials} The data of simulation experiment and illustrative study is available from the corresponding author upon request.

\noindent
\textbf{Code availability} The code of data simulations and illustrative study is available from the corresponding author upon request.

\noindent
\textbf{Authors' contributions} CB conceived the study. CB and HG supervised the study. LZ performed simulations and statistical analysis. TF conceived the illustrative study and provided the data. LZ, CB and HG interpreted statistical results and wrote the manuscript. LZ, TF, CB and HG read and approved the final manuscript.

\comm{

\begin{itemize}
\item Funding
\item Competing interests
\item Ethics approval
\item Consent to participate
\item Consent for publication
\item Availability of data and materials
\item Code availability
\item Authors' contributions
\end{itemize}

\noindent
If any of the sections are not relevant to your manuscript, please include the heading and write `Not applicable' for that section.

\bigskip
\begin{flushleft}%
Editorial Policies for:

\bigskip\noindent
Springer journals and proceedings: \url{https://www.springer.com/gp/editorial-policies}

\bigskip\noindent
Nature Portfolio journals: \url{https://www.nature.com/nature-research/editorial-policies}

\bigskip\noindent
\textit{Scientific Reports}: \url{https://www.nature.com/srep/journal-policies/editorial-policies}

\bigskip\noindent
BMC journals: \url{https://www.biomedcentral.com/getpublished/editorial-policies}
\end{flushleft}

\begin{appendices}

\section{Section title of first appendix}\label{secA1}

An appendix contains supplementary information that is not an essential part of the text itself but which may be helpful in providing a more comprehensive understanding of the research problem or it is information that is too cumbersome to be included in the body of the paper.




\end{appendices}

}


\bibliography{bibliography}


\comm{

}

\end{document}